\documentclass[12pt]{article}
\usepackage{graphicx}
      \usepackage{bm}

\usepackage{multirow}
\usepackage{ctable}
\usepackage{booktabs}
\usepackage{array}
\usepackage{tabularx}
\usepackage{xcolor}
\usepackage{pstricks}
\DeclareMathAlphabet{\mathpzc}{OT1}{pzc}{m}{it}

\newcommand\pubnumber{}
\newcommand\pubdate{\today}

\def\napoli{Institute of Nuclear
Physics, Polish Academy of Sciences, Radzikowskiego 152, PL-31-342 Krak{\'o}w, Poland}
\def\support{\footnote{This study was partially
supported by the Polish National Science Center grant
DEC-2014/15/B/ST2/02528 and by the Center for Innovation and
Transfer of Natural Sciences and Engineering Knowledge in
Rzesz{\'o}w.}}

\def\Title#1{\begin{center} {\Large #1 } \end{center}}
\def\Author#1{\begin{center}{ \sc #1} \end{center}}
\def\Address#1{\begin{center}{ \it #1} \end{center}}

\newcommand\pubblock{\rightline{\begin{tabular}{l} \pubnumber\\
         \pubdate  \end{tabular}}}
\newenvironment{Abstract}{\begin{quotation}  }{\end{quotation}}
\newenvironment{Presented}{\begin{quotation} \begin{center} 
             PRESENTED AT\end{center}\bigskip 
      \begin{center}\begin{large}}{\end{large}\end{center} \end{quotation}}

\def\beq{\begin{equation}}
\def\eeq#1{\label{#1}\end{equation}}
\def\eeqn{\end{equation}}

\def\beqa{\begin{eqnarray}}
\def\eeqa#1{\label{#1}\end{eqnarray}}
\def\eeqan{\end{eqnarray}}

\let\bar=\overbar

\def\L{{\cal L}}

\def\Dslash{\not{\hbox{\kern-4pt $D$}}}
\def\dslash{\not{\hbox{\kern-2pt $\del$}}}

\def\msb{{\bar{\ssstyle M \kern -1pt S}}}

\begin{document}
\begin{titlepage}
\pubblock

\vfill
\Title{Associated production of $D$-mesons with jets at the LHC}
\vfill
\Author{ Rafa{\l} Maciu{\l}a and Antoni Szczurek\support}
\Address{\napoli}
\vfill
\begin{Abstract}
We present several differential distributions for the associated production of charm and dijets.
Both single-parton scattering (SPS) and double-parton scattering (DPS) contributions are calculated
in the $k_T$-factorization approach. We have found regions of 
the phase space where the SPS contribution 
is negligible compared to the DPS one.
The distribution in transverse momentum of
charmed mesons as well as azimuthal correlations ($D^0\mathrm{\textit{-jet}}$, $D^0 \overline{D^0}$) can be used for 
experimental identification of the DPS effects.
\end{Abstract}
\vfill
\begin{Presented}
The 17th Conference on Elastic and Diffractive Scattering, EDS Blois 2017,
26th - 30th June 2017, Prague, Czech Republic
\end{Presented}
\vfill
\end{titlepage}
\def\thefootnote{\fnsymbol{footnote}}
\setcounter{footnote}{0}

\section{Introduction}

Charm particles, quarks and/or mesons, are produced abundantly in double- \cite{Luszczak:2011zp,Maciula:2013kd,Cazaroto:2013fua,vanHameren:2014ava,Maciula:2016wci}
or multiple- \cite{Maciula:2017meb,Cazaroto:2013fua} parton scattering. The cross section for double 
charm  production was shown to grow considerably with the collision
energy \cite{Luszczak:2011zp}. In our previous papers we have explained that the LHCb double charm data \cite{Aaij:2012dz} cannot be explained without inclusion of double-parton scattering. 
Many processes in association with charm quarks or mesons are possible
and can be studied at the LHC. Recently we discussed inclusive production
of single jet associated with $c \bar c$ or charmed mesons \cite{Maciula:2016kkx}.
Quite large cross sections were found there. This reaction was discussed in the presented talk, however will be not
considered in the following due to the limited form of the proceedings contribution. 

Here we discuss inclusive production of dijets in association with $c \bar c$ production. 
We wish to include both single parton scattering (SPS) and 
double parton scattering (DPS) mechanisms and check whether the process can be used to extract the
so-called $\sigma_{eff}$ parameter which governs the strength of double
parton scattering. Therefore we need to focus on how to disantangle single and double parton scattering
contributions for a simultaneous production of $c \bar c$ (or charmed
mesons) and dijets.

\section{Formalism}

\subsection{Single-parton scattering}

Within the $k_T$-factorization approach \cite{kTfactorization} the SPS cross section for 
$pp \to c\bar c + \mathrm{2jets}\, X$ reaction can be written as
\begin{equation}
d \sigma_{p p \to c\bar c + \mathrm{2jets}} = \sum_{ij}
\int d x_1 \frac{d^2 k_{1t}}{\pi} d x_2 \frac{d^2 k_{2t}}{\pi}
{\cal F}_{i}(x_1,k_{1t}^2,\mu^2) {\cal F}_{j}(x_2,k_{2t}^2,\mu^2)
d {\hat \sigma}_{ij \to c \bar c + \mathrm{2part.} }
\; .
\label{cs_formula}
\end{equation}
In the formula above ${\cal F}_{i}(x,k_t^2,\mu^2)$ is a unintegrated
parton distribution function (uPDF) for a given type of parton $i = g, u, d, s, \bar u, \bar d, \bar s$. The uPDFs depend on longitudinal momentum fraction $x$, transverse momentum squared $k_t^2$ of the partons entering the hard process,
and in general also on a (factorization) scale of the hard process $\mu^2$.
The elementary cross section in Eq.~(\ref{cs_formula}) can be written
somewhat formally as:
\begin{equation}
d {\hat \sigma}_{ij \to c \bar c + \mathrm{2part.} } =
\prod_{l=1}^{4}
\frac{d^3 p_l}{(2 \pi)^3 2 E_l} 
(2 \pi)^4 \delta^{4}(\sum_{l=1}^{4} p_l - k_1 - k_2) \times\frac{1}{\mathrm{flux}} \overline{|{\cal M}_{i^* j^* \to c \bar c + \mathrm{2part.}}(k_{1},k_{2})|^2}
\; ,
\label{elementary_cs}
\end{equation}
where $E_{l}$ and $p_{l}$ are energies and momenta of final state particles. Above only dependence of the matrix element on four-vectors of incident partons $k_1$ and $k_2$ is made explicit. In general all four-momenta associated with partonic legs enter.
The matrix element takes into account that both partons entering the hard
process are off-shell with virtualities $k_1^2 = -k_{1t}^2$ and $k_2^2 = -k_{2t}^2$.
We take into account all 9 channels of the $2 \to 4$ type contributing to the cross section at the parton-level:\\
\begin{center}
$\#1 = g \; g \to g \; g \; c \; \bar{c}$ $\;\;\;\;\;\;$ $\#2 = g \; g \to q \; \bar{q} \; c \; \bar{c}$
$\;\;\;\;\;\;$ $\#3 = g \; q \to g \; q \; c \; \bar{c}$\\
$\#4 = q \; g \to q \; g \; c \; \bar{c}$ $\;\;\;\;\;\;$ $\#5 = q \; \bar{q} \to q' \; \bar{q}' \; c \; \bar{c}$
$\;\;\;\;\;\;$ $\#6 = q \; \bar{q} \to g \; g \; c \; \bar{c}$\\
$\#7 = q \; q \to q \; q \; c \; \bar{c}$ $\;\;\;\;\;\;$ $\#8 = q \; q' \to q \; q' \; c \; \bar{c}$
$\;\;\;\;\;\;$ $\#9 = q \; \bar{q} \to q \; \bar{q} \; c \; \bar{c}$.
\end{center}

The calculation has been performed with the help of KaTie \cite{vanHameren:2016kkz}, which is a complete Monte Carlo parton-level event generator for hadron scattering processes. It can can be applied to any arbitrary processes within the Standard Model, for many final-state particles, and for any initial-state partons on-shell or off-shell.

\subsection{Double-parton scattering}

According to the general form of the multiple-parton scattering theory (see \textit{e.g.} Refs.~\cite{Diehl:2011tt,Diehl:2011yj})
the DPS cross sections can be expressed in terms of the double parton distribution functions (dPDFs).
These objects should fulfill sum rules and take into account all the correlations between the two partons. The theory of dPDFs is well established but still not fully applicable for phenomenological studies.
Instead of the general form, one usually follows the assumption of the factorization of the DPS cross section.
Within this framework, the differential DPS cross section for $pp \to c\bar c + \mathrm{2jets}\; X$ reaction can be expressed as follows: 
\begin{equation}
\frac{d\sigma^{DPS}(c \bar c + \mathrm{2jets})}{d\xi_{1}d\xi_{2}} = \sum_{i,j} \;  \frac{1}{\sigma_{eff}} \cdot \frac{d\sigma^{SPS}(g g \to c \bar c)}{d\xi_{1}} \! \cdot \! \frac{\sigma^{SPS}(i j \to \mathrm{2jets})}{d\xi_{2}},
\label{basic_formula}
\end{equation}
where $\xi_{1}$ and $\xi_{2}$ stand for generic phase space kinematical variables for the first and second scattering, respectively.
When integrating over kinematical variables one recovers the commonly used pocket-formula:
\begin{equation}
\sigma^{DPS}(c \bar c + \mathrm{2jets}) = \sum_{i,j} \;  
\frac{\sigma^{SPS}(g g \to c \bar c) \! \cdot \! \sigma^{SPS}(i j \to \mathrm{2jets})}{\sigma_{eff}}\; .
\label{basic_formula}
\end{equation}

The effective cross section $\sigma_{eff}$ provides a proper normalization of the DPS cross section and can be roughly interpreted 
as a measure of the transverse correlation of the two partons inside 
the hadrons. The longitudinal parton-parton correlations should become far less
important as the energy of the collision is increased, due to the increase in the parton multiplicity. It is belived that for small-$x$ partons and for low and intermediate scales the possible longitudinal correlations can be safely
neglected (see \textit{e.g.} Ref.~\cite{Gaunt:2009re}). 
In this paper we use world-average value of $\sigma_{eff} = 15$ mb provided by 
several experiments (see e.g. Refs.~\cite{Astalos:2015ivw,Proceedings:2016tff}).

The cross sections for each step of the DPS mechanism are calculated in the
$k_T$-factorization approach, that is:
\begin{eqnarray}
\frac{d \sigma^{SPS}(p p \to c \bar c \; X_1)}{d y_1 d y_2 d^2 p_{1,t} d^2 p_{2,t}} 
&& = \frac{1}{16 \pi^2 {\hat s}^2} \int \frac{d^2 k_{1t}}{\pi} \frac{d^2 k_{2t}}{\pi} \overline{|{\cal M}_{g^{*} g^{*} \rightarrow c \bar{c}}|^2} \nonumber \\
&& \times \;\; \delta^2 \left( \vec{k}_{1t} + \vec{k}_{2t} - \vec{p}_{1t} - \vec{p}_{2t}
\right)
{\cal F}_{g}(x_1,k_{1t}^2,\mu^2) {\cal F}_{g}(x_2,k_{2t}^2,\mu^2),
\nonumber
\end{eqnarray}
\begin{eqnarray}
\frac{d \sigma^{SPS}(p p \to \mathrm{2jets} \; X_2)}{d y_3 d y_4 d^2 p_{3,t} d^2 p_{4,t}} 
&& = \frac{1}{16 \pi^2 {\hat s}^2} \sum_{ij} \int \frac{d^2 k_{3t}}{\pi} \frac{d^2 k_{4t}}{\pi} \overline{|{\cal M}_{i^{*} j^{*} \rightarrow \mathrm{2part.}}|^2} \nonumber \\
&&\times \;\; \delta^2 \left( \vec{k}_{3t} + \vec{k}_{4t} - \vec{p}_{3t} - \vec{p}_{4t}
\right)
{\cal F}_{i}(x_3,k_{3t}^2,\mu^2) {\cal F}_{j}(x_4,k_{4t}^2,\mu^2). \nonumber \\
\end{eqnarray}
The numerical calculations for both SPS mechanisms are also done within the KaTie code.

\section{Numerical results}

\subsection{$\bm{D^{0} + \mathrm{2jets}}$}

We start with the predictions for single $D^{0}$ meson production in association with exactly two jets.
In this analysis, the $D^{0}$ meson is required to have $|y^{D^{0}}| < 2.5$ and $p_{T}^{D^{0}} > 3.5$ GeV and the rapidities of both associated jets are $|y^{jet}| < 4.9$, which corresponds to the ATLAS detector acceptance. In Table~\ref{tab:cross sections_D} we collect the corresponding integrated cross sections for inclusive $D^{0}+\mathrm{2 jets}$ production in $pp$-scattering at $\sqrt{s} =$ 13 TeV for different cuts on transverse momenta of the associated jets, specified in the left column. We found large cross sections, of the order of a few, and up to even tens of microbarns, depending on the cuts on transverse momenta of the associated jets. The cross sections are dominated by the DPS mechanism with the relative DPS contribution at the level of $70 - 80 \%$.  

\begin{table}[tb]%
\caption{The calculated cross sections in microbarns for inclusive
  $D^{0}+\mathrm{2 jets}$ production in $pp$-scattering at $\sqrt{s} =$ 13 TeV for different cuts on transverse momenta of the associated jets. Here, the $D^{0}$ meson is required to have $|y^{D^{0}}| < 2.5$ and $p_{T}^{D^{0}} > 3.5$ GeV and the rapidities of the both associated jets are $|y^{jet}| < 4.9$, which corresponds to the ATLAS detector acceptance. }

\label{tab:cross sections_D}
\centering %
\begin{tabularx}{1.0\linewidth}{c c c c}
\\[-1.ex] 
\toprule[0.1em] %
\\[-3.ex] 

\multirow{1}{7.5cm}{experimental jet-$p_{T}$ mode} & \multirow{1}{2.cm}{SPS} & \multirow{1}{2.cm}{DPS}  & \multirow{1}{2.cm}{$\frac{DPS}{SPS+DPS}$}  \\ [+0.1ex]
\bottomrule[0.1em]
\multirow{1}{7.5cm}{both jets $p_{T} > 20$ GeV} &                         \multirow{1}{2.cm}{3.74} & \multirow{1}{2.cm}{18.49} & \multirow{1}{2.cm}{$\;\;\;\;$83 \%}  \\ [-0.2ex]
\multirow{1}{7.5cm}{$p_{T}^{lead} > 35$ GeV, $\; p_{T}^{sub} > 20$ GeV} & \multirow{1}{2.cm}{1.76} & \multirow{1}{2.cm}{4.52} & \multirow{1}{2.cm}{$\;\;\;\;$72 \%} \\ [-0.2ex]
\multirow{1}{7.5cm}{$p_{T}^{lead} > 50$ GeV, $\; p_{T}^{sub} > 35$ GeV} & \multirow{1}{2.cm}{0.43} & \multirow{1}{2.cm}{1.25} & \multirow{1}{2.cm}{$\;\;\;\;$74 \%} \\ [-0.2ex]

\hline

\bottomrule[0.1em]

\end{tabularx}
\end{table}

\begin{figure}[!h]
\begin{minipage}{0.47\textwidth}
 \centerline{\includegraphics[width=1.0\textwidth]{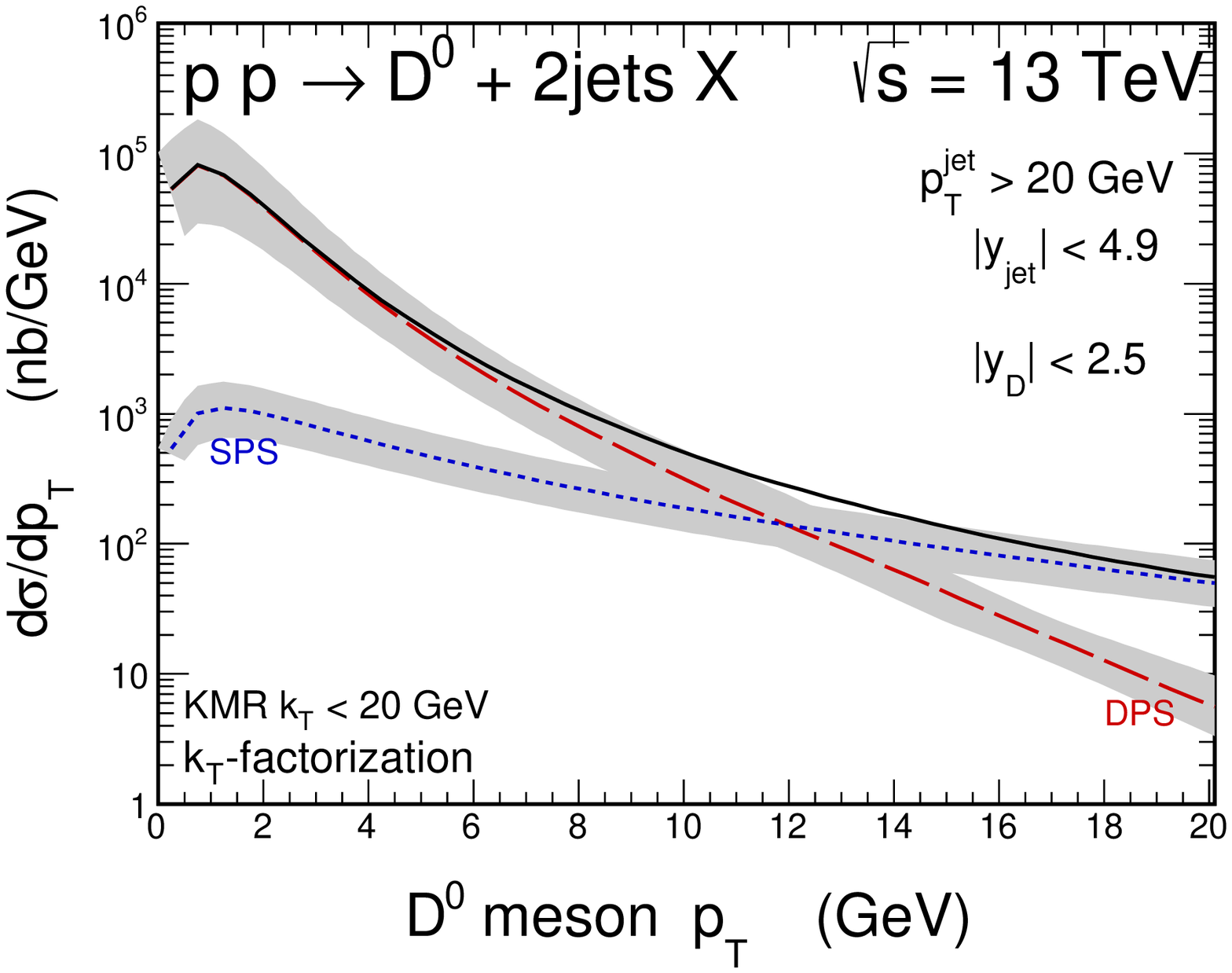}}
\end{minipage}
\hspace{0.5cm}
\begin{minipage}{0.47\textwidth}
 \centerline{\includegraphics[width=1.0\textwidth]{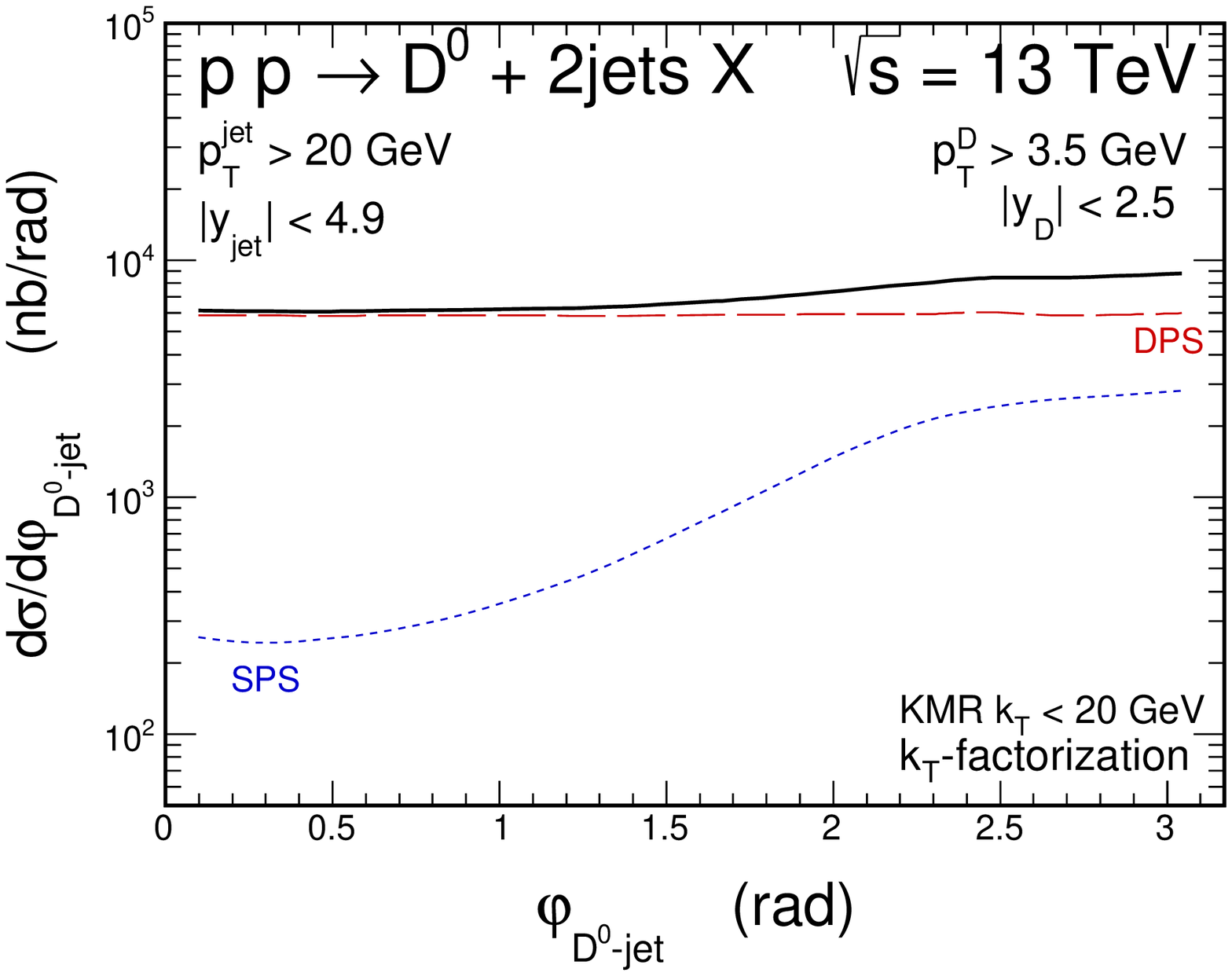}}
\end{minipage}
   \caption{
\small The transverse momentum (left) and azimuthal angle $\varphi_{D^{0}\mathrm{\textit{-jet}}}$ (right) distribution of the $D^{0}$ meson for SPS (dotted) and DPS (dashed) mechanisms for the ATLAS detector acceptance. The solid line represents a sum of the two components. Details are specified in the figure. For example in the left panel we show explicitly theoretical uncertainties, see discussion in the text.  
 }
 \label{fig:ptD}
\end{figure}

In Fig.~\ref{fig:ptD} we show the differential cross section as a function of transverse momenta of the $D^{0}$ meson (left panel) and as a function of the azimuthal angle $\varphi_{D^{0}\mathrm{\textit{-jet}}}$ between the $D^{0}$ meson ($\overline{D^{0}}$ antimeson) and the leading jet (right panel). The DPS (dashed line) and the SPS (dotted line) components are shown separately together with their sum (solid line). We observe that in the region of $D^{0}$ meson transverse momenta $p_{T} < 10$ GeV the DPS mechanism significantly dominates over the SPS one.
We also see that the presence and the dominant role of the DPS component leads to a significant enhancement of the cross section and to a visible decorrelation of the azimuthal distribution in contrast to the pure SPS-based predictions.
In the left panel we plot in addition the typical uncertainty bands of the pQCD calculations for both, the SPS and the DPS components. The shaded bands represent the uncertainties related to the choice of renormalization/factorization scales and charm quark mass, summed in quadrature. We vary the charm quark mass $m_{c} = 1.5 \pm 0.25$ GeV and the scales $\mu^{2}$ by a factor 2, which is a rather standard procedure. The calculated uncertainties are about $\pm 45 \%$ for the SPS and $\pm 65 \%$ for the DPS mechanism. These levels of uncertainty also apply for the integrated cross sections.

\subsection{$\bm{D^{0}\bar{D^{0}} + \mathrm{2jets}}$}

Now we also consider the case of production of the $D^{0}\overline{D^{0}}$-pair in association with two jets.
Both, $D^{0}$-meson and $\overline{D^{0}}$-antimeson are required to enter the ATLAS detector acceptance.
The corresponding theoretical cross sections are collected in Table~\ref{tab:cross sections_DD}. Here, the predicted cross sections for  $D^{0}\overline{D^{0}}+\mathrm{2 jets}$ are slightly
smaller than in the case of $D^{0}+\mathrm{2 jets}$ production (see Table~\ref{tab:cross sections_D}) but still large (in the best scenario, of the order of a few microbarns).
Also the relative DPS contribution is somewhat reduced and varies at the level of $50 - 70 \%$.   

\begin{table}[tb]%
\caption{The same as in Table~\ref{tab:cross sections_D} but for inclusive $D^{0}\overline{D^{0}}+\mathrm{2 jets}$ production. Here both, $D^{0}$ meson and $\overline{D^{0}}$ antimeson are required to enter the ATLAS detector acceptance.}

\label{tab:cross sections_DD}
\centering %
\begin{tabularx}{1.\linewidth}{c c c c}
\\[-1.ex] 
\toprule[0.1em] %
\\[-3.ex] 

\multirow{1}{7.5cm}{experimental jet-$p_{T}$ mode} & \multirow{1}{2.cm}{SPS} & \multirow{1}{2.cm}{DPS}  & \multirow{1}{2.cm}{$\frac{DPS}{SPS+DPS}$}  \\ [+0.1ex]
\bottomrule[0.1em]
\multirow{1}{7.5cm}{both jets $p_{T} > 20$ GeV} &                         \multirow{1}{2.cm}{1.10} & \multirow{1}{2.cm}{2.35} & \multirow{1}{2.cm}{$\;\;\;\;$68 \%}  \\ [-0.2ex]
\multirow{1}{7.5cm}{$p_{T}^{lead} > 35$ GeV, $\; p_{T}^{sub} > 20$ GeV} & \multirow{1}{2.cm}{0.55} & \multirow{1}{2.cm}{0.58} & \multirow{1}{2.cm}{$\;\;\;\;$51 \%} \\ [-0.2ex]
\multirow{1}{7.5cm}{$p_{T}^{lead} > 50$ GeV, $\; p_{T}^{sub} > 35$ GeV} & \multirow{1}{2.cm}{0.15} & \multirow{1}{2.cm}{0.14} & \multirow{1}{2.cm}{$\;\;\;\;$52 \%} \\ [-0.2ex]

\hline

\bottomrule[0.1em]

\end{tabularx}
\end{table}

\begin{figure}[!h]
\begin{center}
\begin{minipage}{0.47\textwidth}
 \centerline{\includegraphics[width=1.0\textwidth]{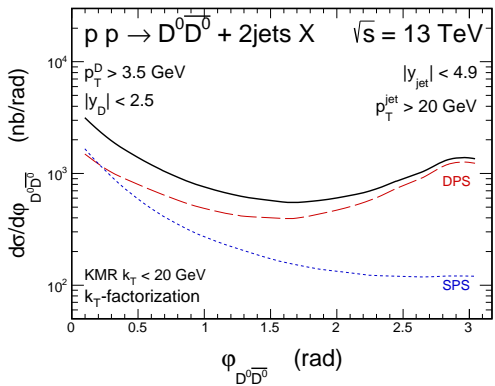}}
\end{minipage}
   \caption{
\small The azimuthal angle $\varphi_{D^{0}\overline{D^{0}}}$ distribution for SPS (dotted) and DPS (dashed) mechanisms for the ATLAS detector acceptance. The solid line represents a sum of the two components. Details are specified in the figure.}
 \label{fig:phiDDbar}
 \end{center}
\end{figure}

In the case of the $D^{0}\overline{D^{0}}+\mathrm{2 jets}$ final state we also find a very interesting correlation observable that may be useful to distinguish between the DPS and SPS mechanisms. 
Figure~\ref{fig:phiDDbar} presents the distributions in azimuthal angle $\varphi_{D^{0}\overline{D^{0}}}$
between the $D^{0}$ meson and $\overline{D^{0}}$ antimeson in the case of $D^{0}\overline{D^{0}}+\mathrm{2 jets}$ production.
One can observe an evident enhancement of the cross section in the region of $\varphi_{D^{0}\overline{D^{0}}} > \frac{\pi}{2}$ caused by the presence of the DPS mechanism.

More details can be found in our original paper \cite{Maciula:2017egq}.


\begin{thebibliography}{99}

\bibitem{Luszczak:2011zp} 
  M.~{\L}uszczak, R.~Maciu{\l}a and A.~Szczurek,
  Phys.\ Rev.\ D {\bf 85}, 094034 (2012).

\bibitem{Maciula:2013kd} 
  R.~Maciu{\l}a and A.~Szczurek,
  Phys.\ Rev.\ D {\bf 87}, no. 7, 074039 (2013).

\bibitem{Cazaroto:2013fua} 
  E.~R.~Cazaroto, V.~P.~Goncalves and F.~S.~Navarra,
  Phys.\ Rev.\ D {\bf 88}, no. 3, 034005 (2013).

\bibitem{vanHameren:2014ava} 
  A.~van Hameren, R.~Maciu{\l}a and A.~Szczurek,
  Phys.\ Rev.\ D {\bf 89}, no. 9, 094019 (2014).

\bibitem{Maciula:2016wci} 
  R.~Maciu{\l}a, V.~A.~Saleev, A.~V.~Shipilova and A.~Szczurek,
  Phys.\ Lett.\ B {\bf 758}, 458 (2016).

\bibitem{Maciula:2017meb} 
  R.~Maciu{\l}a and A.~Szczurek,
  arXiv:1703.07163 [hep-ph].


\bibitem{Aaij:2012dz}
  R.~Aaij {\it et al.} [LHCb Collaboration],
  J. High Energy Phys. {\bf 06} (2012) 141;
   Addendum: J. High Energy Phys. {\bf 03} (2014) 108.

\bibitem{Maciula:2016kkx} 
  R.~Maciu{\l}a and A.~Szczurek,
  Phys.\ Rev.\ D {\bf 94}, no. 11, 114037 (2016).


\bibitem{kTfactorization}
S.~Catani, M.~Ciafaloni and F.~Hautmann, Phys. Lett. B242 (1990) 97.

\bibitem{vanHameren:2016kkz} 
  A.~van Hameren,
  arXiv:1611.00680 [hep-ph].


\bibitem{Diehl:2011tt}
  M.~Diehl and A.~Schafer,
  Phys.\ Lett.\ B {\bf 698} (2011) 389
  [arXiv:1102.3081 [hep-ph]].
  
\bibitem{Diehl:2011yj} 
  M.~Diehl, D.~Ostermeier and A.~Schafer,
  JHEP {\bf 1203}, 089 (2012)
  Erratum: [JHEP {\bf 1603}, 001 (2016)]
  [arXiv:1111.0910 [hep-ph]].

\bibitem{Gaunt:2009re} 
  J.~R.~Gaunt and W.~J.~Stirling,
  JHEP {\bf 1003}, 005 (2010)
  [arXiv:0910.4347 [hep-ph]].
  
\bibitem{Astalos:2015ivw} 
  R.~Astalos {\it et al.},
  "Proceedings of the Sixth International Workshop on Multiple Partonic Interactions at the Large Hadron Collider",
  arXiv:1506.05829 [hep-ph].

\bibitem{Proceedings:2016tff} 
  H.~Jung, D.~Treleani, M.~Strikman and N.~van Buuren,
  "Proceedings, 7th International Workshop on Multiple Partonic Interactions at the LHC (MPI@LHC 2015) : Miramare, Trieste, Italy, November 23-27, 2015",
  DESY-PROC-2016-01.

\bibitem{Maciula:2017egq} 
  R.~Maciu{\l}a and A.~Szczurek,
  Phys.\ Rev.\ D {\bf 96}, no. 7, 074013 (2017).

\end{thebibliography}
\end{document}